\documentclass{Interspeech}

\interspeechcameraready

\title{Enhancing Lyrics Transcription on Music Mixtures with Consistency Loss}

\author[affiliation={1,*}]{Jiawen}{Huang}
\author[affiliation={2}]{Felipe}{Sousa}
\author[affiliation={2}]{Emir}{Demirel}
\author[affiliation={1}]{Emmanouil}{Benetos}
\author[affiliation={2}]{Igor}{Gadelha}

\affiliation{Centre for Digital Music}{Queen Mary University of London}{UK}
\affiliation{}{Music.AI}{USA}
\email{jiawen.huang@qmul.ac.uk, felipe.sousa@music.ai, emir.demirel@music.ai, emmanouil.benetos@qmul.ac.uk, igor@music.ai}


\usepackage{booktabs}
\usepackage{svg}
\usepackage{amsmath}
\usepackage{amssymb}

\usepackage[normalem]{ulem}
\usepackage{arydshln}
\usepackage{multirow}

\begin{document}

\maketitle

\begin{abstract}
Automatic Lyrics Transcription (ALT) aims to recognize lyrics from singing voices, similar to Automatic Speech Recognition (ASR) for spoken language, but faces added complexity due to domain-specific properties of the singing voice. While foundation ASR models show robustness in various speech tasks, their performance degrades on singing voice, especially in the presence of musical accompaniment. This work focuses on this performance gap and explores Low-Rank Adaptation (LoRA) for ALT, investigating both single-domain and dual-domain fine-tuning strategies. We propose using a consistency loss to better align vocal and mixture encoder representations, improving transcription on mixture without relying on singing voice separation. Our results show that while na\"{i}ve dual-domain fine-tuning underperforms, structured training with consistency loss yields modest but consistent gains, demonstrating the potential of adapting ASR foundation models for music.

\end{abstract}

\begingroup
\renewcommand\thefootnote{}\footnotetext{* The author performed this work as an intern at Music.AI.}
\addtocounter{footnote}{0}
\endgroup

\section{Introduction}
Automatic Lyrics Transcription (ALT) is the task of recognizing the linguistic content in singing voice recordings. 
It is the equivalent of automatic speech recognition (ASR) for speech, but presents additional challenges due to the unique characteristics of singing voice and the interference caused by music accompaniment.
Despite significant challenges, ALT persevered as an active research field within music information retrieval research for over multiple decades \cite{DBLP:journals/ejasmp/MesarosV10}, and it has various applications in the music industry ranging from enhancing search capabilities in music archives and interactive applications such as music education and karaoke systems.

Previous research often takes on the ALT task within multiple acoustic domains based on the presence of music accompaniment. Music accompaniment, and similarly backing vocals introduce additional information within the acoustic scene, which can make it difficult for the model to focus on the lyrics and the singing voice. Many ALT studies have explored ways to mitigate this challenge, such as training on both music mixtures and their vocal-separated versions \cite{DBLP:conf/ismir/DemirelAD21}. Other approaches, like joint training with source separation \cite{DBLP:journals/taslp/GaoGL23}, have been proposed to address this issue.

As foundation ASR models like WavLM \cite{DBLP:journals/jstsp/ChenWCWLCLKYXWZ22}, Wav2Vec2 \cite{DBLP:conf/nips/BaevskiZMA20}, and Whisper \cite{DBLP:conf/icml/RadfordKXBMS23} have shown impressive performance in noisy and varying acoustic conditions, ALT researchers have shifted focus to adapting these models to the singing domain  \cite{DBLP:conf/icassp/WangWLJ24, DBLP:conf/ismir/OuGW22}. To adapt these models to the music domain, fine-tuning is required. However, full fine-tuning can be computationally expensive and may be prone to overfitting, especially with smaller train datasets. Parameter-efficient fine-tuning (PEFT) techniques, such as Low-Rank Adaptation (LoRA) \cite{DBLP:conf/iclr/HuSWALWWC22}, address this challenge by updating only a small subset of model parameters while keeping the majority frozen. PEFT on foundation models has shown promising results in various applications \cite{han2024parameterefficient}, making it a suitable approach for the ALT task which is a research domain with limited well-annotated data availability.

\begin{figure}[t!]
  \centering
  \includegraphics[width=\columnwidth]{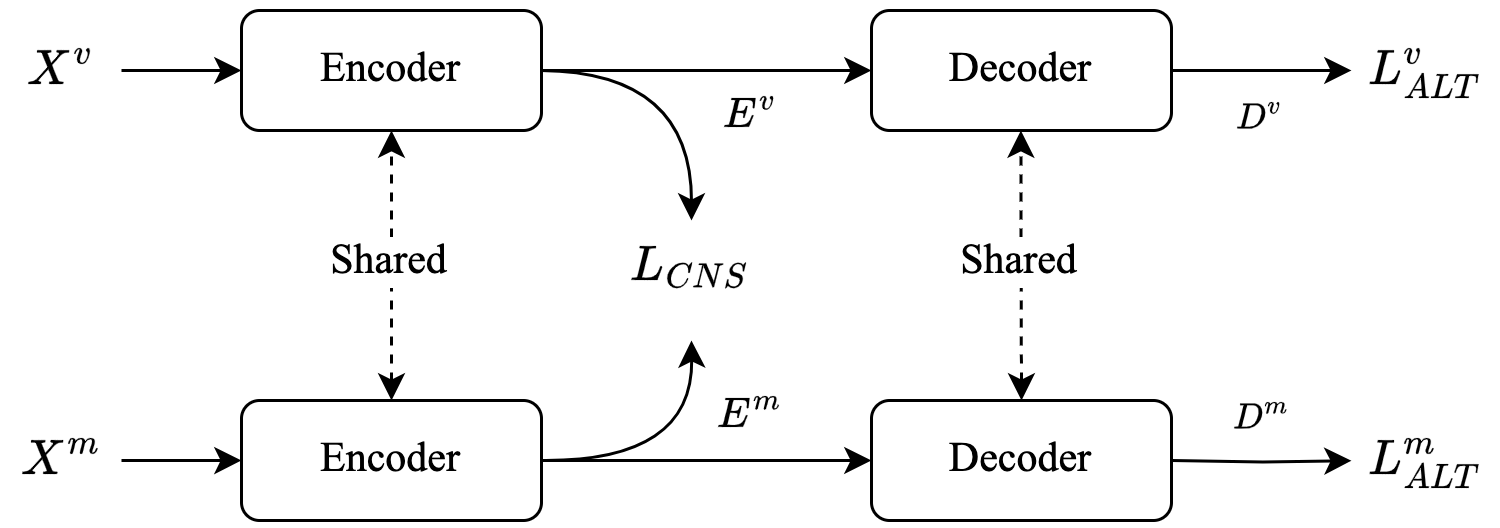}
  \caption{Diagram of dual-domain fine-tuning with consistency loss.}
  \label{fig:diagram}
\end{figure}

Many recent successful ALT systems still rely on singing voice separation (SVS) as preprocessing for improved performance \cite{DBLP:conf/ismir/OuGW22}. In this approach, ALT performance is affected by the SVS model's vocal separation quality, which poses a challenge in optimizing the overall ALT pipelines as SVS models are trained separately. Moreover, SVS adds additional computational complexity during inference, making it less ideal for real-world applications. To reduce the reliance on SVS for improved performance, we aim to handle music mixtures directly by treating background music as noise. This requires the model to learn vocal representations that are invariant to music accompaniment. However, achieving this alignment is challenging due to the mismatch between these two domains, which can hinder performance generalization. 

In this work, we explore parameter-efficient fine-tuning (specifically, Low-Rank Adaptation) of a speech foundation model for ALT. We investigate single and dual-domain fine-tuning strategies, where models are trained on either with or without source separated vocals. Our results highlight the challenge of optimizing dual-domain training, as na\"{i}ve approaches underperform compared to vocal-only fine-tuning. To address this, we propose using a consistency loss to better align vocal and mixture representations. While the improvement is modest, it demonstrates the potential of structured dual-domain training. To promote reproducibility, we publicly share the annotations of the datasets used in experiments. Additionally, we release new annotated data for benchmarking ALT in two new languages, Italian and Portuguese~\footnote{The released data are available at: \url{https://github.com/weAreMusicAI/alt-datasets-interspeech2025}}.

\begin{table}[t]
    \centering
    \begin{tabular}{lcc}
        \toprule
        \textbf{WER} & \textbf{English} & \textbf{Overall} \\
        \midrule
        No SVS & 36.80 & 35.42 \\
        HTDemucs SVS    & 37.50 & 39.09\\
        Music.AI SVS    & 32.36 & 31.86 \\
        \bottomrule
    \end{tabular}
    \caption{A comparison of Whisper \texttt{large-v2} performance on mixtures, HTDemucs-separated vocals, and vocals separated by our internal Music.AI SVS model.}
    \label{tab:wer_svs}
\end{table}

\section{Preliminaries}

\subsection{PEFT on foundation models}

As ASR models get bigger, they require more and more computational resources which often may not be available for researchers. To address the challenges of fine-tuning very large-scale transformers, parameter-efficient fine-tuning (PEFT) techniques have become popular. PEFT reduces the number of trainable parameters, thereby decreasing the computational cost by limiting the number of gradients that need to be computed. This is usually achieved through add-on adapter layers. One such popular method is Low-Rank Adaptation (LoRA) \cite{DBLP:conf/iclr/HuSWALWWC22}. This technique has been shown to effectively fine-tune foundation models for tasks such as multilingual ASR and emotion recognition \cite{prasad24_interspeech, 10446645}. By reducing the need to train the entire model, LoRA significantly lowers the computational demands while still achieving strong performance.

\subsection{Whisper for ALT with SVS}
Whisper is particularly known for its resilience to various noise conditions, including music. Previous studies have explored its application for ALT \cite{DBLP:conf/ismir/ZhuoYPMLZLDFLBC23, cifka-2024-jam-alt}. Its performance on ALT can vary depending on the singing voice separation (SVS) pre-processing tool used. In Jam-ALT \cite{cifka-2024-jam-alt} the authors studied the impact of a specific vocal separation model (HTDemucs \cite{DBLP:conf/icassp/RouardMD23}) on the two Whisper \texttt{large} models (\texttt{v2} and \texttt{v3}). Both \texttt{v2} and \texttt{v3} perform worse after applying vocal separation. However, our preliminary experiments show that it is possible to achieve better results after vocal separation. Table~\ref{tab:wer_svs} lists the word error rates on Multi-Lang Jamendo \cite{10096725} on mixtures, HTDemucs-separated vocals, and our internal Music.AI SVS model. The dataset and evaluation is discussed in Section~\ref{sec:exp}.

\subsection{Consistency loss}

In the recent literature, dual-path adaptive training is proposed as an alternative training framework for noise-robust ASR. According to these approaches, an auxiliary loss is included as penalty which is computed as the distance between the intermediary representations of encoder layers. We refer to this auxiliary loss as the consistency loss. Shi and Kawahara \cite{shi2024dual} proposed this in the feature extraction module to handle data mismatch between pretraining and evaluation, where a mean squared error (MSE) is used as the consistency loss between the noisy and the clean features. In \cite{DBLP:conf/icassp/ZhuZZWFD22}, an MSE loss between clean and noisy features are combined with the contrastive loss for Wav2Vec2.0 pretraining.
Beyond feature pretraining, similar techniques have also been applied directly to ASR.
Hu et al. \cite{DBLP:conf/interspeech/HuH0C23} introduced speech enhancement as a front-end to noisy ASR, and proposed a dual-path style learning approach to enhance robustness. Consistency losses are computed both at intermediate encoder layers, and the decoder output.
Recently, Yao et al. \cite{yao2024cr} proposed using a consistency loss between two CTC distributions from different augmented versions of the input speech, using the Kullback-Leibler divergence.

\section{Method}

Due to their large-scale pretraining feature, the encoder representations of the open-source checkpoints for the Whisper models are shown  to capture rich background sound information, making them useful for tasks such as audio tagging \cite{DBLP:conf/interspeech/0001KKG23}. When a singing voice is accompanied by an accompaniment, this additional musical information is also encoded, even though it is irrelevant to the ALT task, which focuses solely on transcribing the singing voice. This suggests potential improvements by refining Whisper’s ability to focus more on vocal features. 

To achieve this, we propose using a consistency loss that encourages the model to produce similar encoded representations for both separated vocals and mixture inputs. The aim is for the model to emphasize the singing voice while minimizing the influence of the accompaniment.

Figure~\ref{fig:diagram} illustrates the dual-path fine-tuning process with consistency loss. Let $X^d, \forall d \in \{v,m\}$ represent the Mel-spectrograms with $v$ and $m$, represent the vocal and the mixture paths of the overall architecture. $y$ represents the tokenized ground truth lyrics. Both inputs are processed by the Whisper encoder, $enc$ and decoder, $dec$. The encoded features produced by the encoder are denoted as $E = enc(X^d)$, while the corresponding decoder outputs are represented as $D = dec(E^d)$.

Following Whisper's standard training procedure, we use the cross-entropy loss, $L_{\text{CE}}$ to compute the ALT losses based on the path-specific decoder outputs:
\begin{equation}
L_{\text{ALT}}^d = L_{\text{CE}}(D^d, y),
\end{equation}
To encourage the model to learn similar representations for vocal and mixture inputs, we introduce a consistency loss computed between the encoded features:
\begin{equation}
L_{\text{CNS}} = l(E^v, E^m)
\end{equation}
where the $l$  is the loss term which can be either an L1 or L2 loss. The final loss is computed as:
\begin{equation}
    L = \frac{L_{\text{ALT}}^v+L_{\text{ALT}}^m}{2} + w L_{\text{CNS}}
\end{equation}
where $w$ is a tunable parameter that balances the ALT loss and consistency loss. We conduct a series of experiments to determine the best loss function and the weight $w$.

\section{Experiments}\label{sec:exp}

\subsection{Datasets}

\begin{table}[t!]
    \centering
    \begin{tabular}{l l r r}
        \toprule
        \textbf{Dataset} & \textbf{Split} & \textbf{Songs} & \textbf{30s Segments} \\
        \midrule
        DALI v2.0 & train & 6672 & 41451 \\
        DALI v2.0  & dev & 100 & 623 \\
        Muljam v2.0 & train & 5599 & 32393 \\
        Muljam v2.0 & dev & 114 & 682 \\
        Multi-Lang Jamendo & test & 79 (+ 39)& - \\ 
        \bottomrule
    \end{tabular}
    \caption{Dataset Statistics of DALI v2.0, MulJam v2.0, and the Multi-lang Jamendo with the new Italian and Portuguese sets.}
    \label{tab:dataset_statistics}
\end{table}

For training, we use two datasets commonly used in recent ALT research, namely DALI v2.0 \cite{meseguer2020creating} and MulJam v2.0 \cite{DBLP:conf/nips/YuanMLZCYZLHTDW23}. DALI v2.0 is a multimodal, multilingual dataset for polyphonic music retrieved from YouTube, with time-aligned lyrics. To ensure language diversity, we selected the five languages with at least 200 songs each (English, French, Spanish, German, and Italian). MulJam v2.0 is a recently released multilingual lyrics transcription dataset in six languages (the above-mentioned five and Russian), sourced from MTG-Jamendo \cite{bogdanov2019mtg}. The dataset properties are listed in Table~\ref{tab:dataset_statistics}. 

For evaluation, we used the Multi-Lang Jamendo dataset which has 79 songs in 4 languages (English, French, German, Spanish). 
Additionally, we release new evaluation sets in Portuguese (PT) and Italian (IT) languages, containing 20 and 19 songs respectively. These sets are curated following the principles outlined in \cite{cifka-2024-jam-alt}, ensuring consistency with the data properties of the older version. The song selections were sourced from the Jamendo website, associated with proper open-source licensing that would allow sharing for open-science. The lyrics annotations are prepared by three trained annotators who manually corrected identified errors which occasionally involve missing/repeated song sections or typos. Upon completion of the initial annotations, a peer review was conducted, wherein each annotator assessed the accuracy and consistency of the annotations performed by the others. 
In the rest of the paper, ``overall'' refers to the Multi-Lang Jamendo dataset together with the new IT and PT sets.

\subsection{Pre-processing}

Initial inspection on the training sets revealed that the noise in lyrics annotations were non-negligible. To improve data quality, we first remove the over-repeated vowels which represent musical cues embedded in the lyrics. A more significant source of noise was related to inaccurate timestamp annotations and language metadata. To address this, we exploited our internal Music.AI lyrics alignment tool to generate more precise alignments at the lyrics-line level. Prior to generating these alignments, we transcribed the lyrics with an internal Music.AI pretrained model for MulJam v2.0 and relabeled the languages accordingly. 
This results in a total of 38 languages. Notably, English is the dominant language, accounting for 70.27\%, while Italian and Portuguese make up 4\% and 0.27\%, respectively.
Furthermore, tracks without any singing voice content were discarded. For preparing the training data, we merge consecutive lyrics line-level samples to form segments up to 30 seconds. Music.AI SVS is applied to generate separated vocal tracks.
We also share the refined versions of the dataset annotations.

\useunder{\uline}{\ul}{}
\begin{table*}[t!]
\centering
\begin{tabular}{cllllll|llll}
\toprule
\textbf{WER} & \textbf{$L_\text{CNS}$} & $w$ & \multicolumn{1}{c}{\textbf{EN Mix}} & \multicolumn{1}{c}{\textbf{Overall Mix}} & \multicolumn{1}{c}{\textbf{EN Voc}} & \multicolumn{1}{c|}{\textbf{Overall Voc}} & \multicolumn{1}{c}{\textbf{IT Mix}} & \multicolumn{1}{c}{\textbf{PT Mix}} & \multicolumn{1}{c}{\textbf{IT Voc}} & \multicolumn{1}{c}{\textbf{PT Voc}} \\ \midrule
raw whisper  & -            & -          & 36.80                               & 35.42                                    & 35.80                               & 33.49                                    & 35.98                               & 35.53                               & 34.72                               & 32.68                               \\ \hdashline
voc only     & -            & -          & 37.31                               & \textbf{32.85}                           & {\ul \textbf{30.78}}                & {\ul \textbf{29.67}}                     & {\ul \textbf{31.98}}                             & {\ul \textbf{33.09}}                & {\ul \textbf{32.28}}                & {\ul \textbf{31.21}}                \\
mix only     & -            & -          & 37.23                               & 33.67                                    & 35.62                               & 33.67                                    & 33.76                & 36.13                               & 36.73                               & 34.90                               \\ \hdashline
random       & -            & -          & \textbf{34.90}                      & 33.33                                    & 33.20                               & 33.35                                    & 34.15                               & 35.10                               & 37.36                               & 34.01                               \\
both         & -            & -          & 36.67                               & 33.79                                    & 33.40                               & 32.72                                    & 34.48                               & 36.71                               & 35.96                               & 34.37                               \\ \midrule
\multirow{6}{*}{w/CNS} & L1           & 0.1        & 35.75                               & 33.26                                    & 34.31                               & 32.73                                    & 35.11                               & 35.00                               & 35.63                               & 33.24                               \\
                       & L1           & 1.0        & 35.53                               & 32.80                                    & 34.70                               & 32.76                                    & 35.41                               & 34.35                               & \textbf{35.41}                      & 32.54                               \\
                       & L1           & 10.0       & 35.85                               & 32.48                                    & 34.55                               & \textbf{32.21}                           & \textbf{34.09}                      & 34.20                               & 36.32                               & 32.37                               \\
                       & L2           & 0.1        & {\ul \textbf{34.07}}                & 32.94                                    & \textbf{33.70}                      & 32.67                                    & 34.48                               & 35.89                               & 35.82                               & 33.86                               \\
                       & L2           & 1.0        & 34.53                               & {\ul \textbf{32.36}}                     & 34.47                               & 32.47                                    & 35.77                               & 34.44                               & 35.65                               & 32.56                               \\
                       & L2           & 10.0       & 35.80                               & 32.68                                    & 34.41                               & 32.31                                    & 34.86                               & \textbf{34.11}                      & 36.15                               & \textbf{32.32}                     \\ \bottomrule
\end{tabular}
\caption{Word Error Rates (WER) evaluated on both mixture (\textbf{Mix}) and separated vocals (\textbf{Voc}) conditions. The bold text highlights the best performance within each section, whereas underlined bold text indicates the overall best performance. The $L_\text{CNS}$ column specifies the consistency loss function used (L1 or L2), and the $w$ column denotes the corresponding weighting parameter. }
\label{tab:wer_cns}
\end{table*}

\subsection{Training and inference}\label{sec:training_details}

All models are fine-tuned using the ADAM optimizer. The learning rate is set to $5e^{-7}$ for most models, except for \textbf{voc only}, which uses $1e^{-6}$. These values are selected through a parameter search within $[1e^{-7}, 5e^{-7}, 1e^{-6}]$. The learning rate follows a linear warmup and decay schedule, with the first 10\% of training for warmup and the remainder for decay. The batch size is set to 64.
For the parameter-efficient finetuning, we employed the Low-Rank Adaptation (LoRA), with a rank of 8, a dropout rate of 0.5, and  $\alpha = 8$.

We employ a long-form decoding method similar to \cite{DBLP:conf/icml/RadfordKXBMS23} that processes audio with sliding windows of 30 seconds. Prompts were not used as priors during inference, as they did not yield consistent performance improvements as reported in \cite{DBLP:conf/ismir/ZhuoYPMLZLDFLBC23}. The model operates without any indication of whether the input segment is a vocal track or a mixture.

\subsection{Evaluation}

We assess model performance on both mixtures and separated vocals. Both the ground truth and predicted lyrics are normalized by converting to lowercase, removing punctuation, and eliminating consecutive whitespace. Due to space constraints, we report results for English subsets of the Multi-Lang Jamendo, and the newly introduced Portuguese and Italian. We also include the overall WER scores (in Table \ref{tab:wer_cns}) summarized over all languages (i.e. all 118 songs in total).

\subsection{Baselines}

To evaluate the effectiveness of the proposed methods, we compare them against multiple baselines:

\noindent
\textbf{Whisper (no fine-tuning)}: 
The first baseline is Whisper \texttt{large-v2} without any fine-tuning, using the inference pipeline described in Section~\ref{sec:training_details}.

\noindent
\textbf{Single-domain fine-tuning}: 
Vocals only fine-tuning (\textbf{voc only}) trains exclusively on separated vocals, without any mixture input. It serves for comparison with models trained with mixture data.
Mixture only fine-tuning (\textbf{mix only}) trains only on mixture data, directly adapting Whisper to the target domain without seeing separated vocals during training.

\noindent
\textbf{Dual-domain fine-tuning}: 
Random input fine-tuning (\textbf{random}) trains on both vocals and mixture data, with each input randomly selected as either vocals or mixture. This approach allows the model to learn from both data types without structured pairing.
Dual input fine-tuning (\textbf{both}) processes paired vocals and mixture samples, computing the loss as the average of $L_{ALT}^v$ and $L_{ALT}^m$. Unlike \textbf{random}, where inputs are chosen independently, this baseline processes paired vocals and mixture samples during fine-tuning. This setup provides a direct comparison to our proposed approach without incorporating the consistency loss.

\section{Results and discussion}

\subsection{Single-domain fine-tuning}

The top section in Table~\ref{tab:wer_cns} presents the WERs for the baseline models. We can make several observations.
Both \textbf{voc only} and \textbf{mix only} outperform raw Whisper on overall mixture and overall vocals, indicating that fine-tuning on singing voice data is beneficial. However, for the English subset, performance sometimes worsens, suggesting potential language-specific variations.

The \textbf{voc only} model demonstrates strong performance on overall mixture, indicating that fine-tuning only on vocals enhances the model’s ability to generalize to mixture tasks. On the other hand, \textbf{mix only} improves mixture performance but doesn’t perform as well on vocals, showing that training exclusively on mixtures doesn’t sufficiently capture vocal features.

\subsection{Dual-domain fine-tuning}

While the \textbf{random} model benefits from exposure to both data types, it doesn’t outperform \textbf{voc only} on mixture tasks. This suggests that focused fine-tuning on vocals is more effective for mixture transcription than random exposure to both domains. The \textbf{both} model, trained with paired vocals and mixture, performs worse than \textbf{random} and \textbf{voc only}, but better than \textbf{mix only} on overall mixture. This suggests that simply averaging the loss without enforcing consistency constraints does not sufficiently enhance performance on mixture data. Among all fine-tuned models, only \textbf{random} and \textbf{both} show consistent improvement across English and Overall metrics.

\subsection{Dual-domain fine-tuning with consistency Loss}

The bottom section of Table~\ref{tab:wer_cns} compares the WERs for models fine-tuned with and without consistency loss. 
Introducing consistency loss, whether using L1 or L2, consistently improves WERs for both English mix and overall mix compared to the \textbf{both} model without consistency loss. Notably, all consistency loss models outperform all baselines except \textbf{voc only} on overall mix, confirming the effectiveness of the proposed consistency loss approach.

Most models trained with consistency loss also show improvements on overall vocal compared to \textbf{both}. This demonstrates that consistency loss benefits both mixture and vocal performances, although the impact is more pronounced for mixture data. Among the models utilizing consistency loss, L2 loss with a weight of 1.0 achieves the best WER on overall mix, outperforming all other models, including \textbf{voc only}. This highlights the effectiveness of L2 loss in aligning mixture and vocal representations, though other configurations remain competitive.

\subsection{Comparison between single and dual-domain}
The best overall mixture WER is achieved by a consistency loss model, but the improvement over the \textbf{voc only} is marginal. Meanwhile, \textbf{voc only} remains the most effective approach for other metrics, highlighting the importance of strong vocal representations for ALT performance.

Na\"{i}ve dual-domain strategies (\textbf{random} and \textbf{both}), do not show clear advantages over single-domain fine-tuning, suggesting that simply exposing the model to both data types does not effectively bridge the gap between vocals and mixtures. This underscores the challenge of optimizing dual-domain fine-tuning. While incorporating mixture data is intuitively beneficial, it does not easily lead to better generalization. 
One possible reason is that exposure to mixture data may reduce the model's acoustic generalizability, making it more sensitive to music styles. In contrast, \textbf{voc only} focuses solely on vocals, avoiding this issue. Since correlations between language and genre have been observed \cite{patel2010music}, dual-domain fine-tuning may introduce genre-dependent biases, leading to worse performance for underrepresented languages. Further investigation is needed to assess the impact of music accompaniment and to explore strategies for effectively leveraging both domains. Consistency loss provides a slight improvement, marking a step in that direction.

\subsection{Results on Italian and Portuguese}
For Italian and Portuguese, the results indicate that consistency loss does not improve WERs. In fact, the best performances are consistently achieved by the \textbf{voc only} model. Although consistency loss proves beneficial for English and overall mixture, it does not offer the same advantage for Italian and Portuguese. This discrepancy could be attributed to language-specific characteristics, differences in data distribution, or genre-dependent biases. Notably, Italian and Portuguese account for less than 5\% of the training data. These findings suggest that for underrepresented languages, the \textbf{voc only} baseline demonstrates the greatest robustness, consistently achieving the best performance across all conditions.

\section{Conclusion}

In this work, we explored the use of LoRA finetuning on Whisper for improving ALT and proposed a consistency loss for further improvements on music mixtures. Our results show that while na\"{i}ve dual-domain fine-tuning is ineffective, consistency loss provides modest but consistent improvements by better aligning vocal and mixture representations. Among different configurations, L2 loss with a weight of 1.0 yields the best performance on mixtures while maintaining vocal transcription quality. Moreover, we introduced an open-source extension to the Multi-Lang Jamendo dataset with two new languages. Through our experiments we hope to reduce the research gap regarding the use of SVS for the ALT task. 

Future work includes extending this approach to other pretrained models and exploring its effectiveness in self-supervised learning frameworks. Additionally, leveraging larger datasets, including unlabeled singing data, could further enhance model robustness and generalization.


\section{Acknowledgements}

JH is a research student at the UKRI Centre for Doctoral Training in Artificial Intelligence and Music, supported jointly by UK Research and Innovation [grant number EP/S022694/1] and Queen Mary University of London.

The authors would like to thank Vin\'{\i}cius Marques, Angelo Maugeri, Ana Rachel Melo Nascimento, and Ana Lu\'{\i}sa Runze from Music.AI for their help with data annotation and cleaning.

\bibliographystyle{IEEEtran}
\bibliography{mybib}

\end{document}